\documentclass[epj,twocolumn]{webofc}
\usepackage[varg]{txfonts}   
\usepackage{graphicx}

\newcommand{\kmps}{\;{\rm km\;s^{{-}1}}}

\wocname{epj}
\woctitle{Seismology of the Sun and the Distant Stars 2016}
\begin{document}
\title{On possible explanations of pulsations in Maia stars}
%

\author{\firstname{Jadwiga} \lastname{Daszy\'nska-Daszkiewicz}\inst{1}\fnsep\thanks{\email{daszynska@astro.uni.wroc.pl}} \and
        \firstname{Przemys{\l}aw} \lastname{Walczak}\inst{1} \and
        \firstname{Alexey} \lastname{Pamyatnykh}\inst{2}
}

\institute{Inytut Astronomiczny, Uniwersytet Wroc{\l}awski, ul. Kopernika 11, Wroc{\l}aw, Poland
\and
           Nicolaus Copernicus Astronomical Center, Bartycka 18, 00-716, Warsaw, Poland
          }

\abstract{%
The long-time photometric surveys in a few young open clusters allowed to identify
the light variability in stars located on the HR diagram between the well defined $\delta$ Scuti variables
and Slowly Pulsating B-type stars. Several objects of this type were suggested also from
the analysis of the Kepler data. Assuming the pulsational origin of this variability,
we try to explain the observed frequencies with pulsational models involving rotation and/or modification of the mean opacity profile.
}
\maketitle
%
\section{Introduction}
\label{intro}
Pulsating variables between the instability strips of $\delta$ Scuti and Slowly Pulsating B-type (SPB) stars have been suggested many years ago by Struve \cite{1955S&T....14..492S}.
Although later a member of the Pleiades Maia appeared to be a constant star (Struve \citep{1957ApJ...125..115S}), the idea remained and the search have continued for many years
(McNamara \citep{1985ApJ...289..213M}, Scholz \citep{1998A&A...337..447S}).
Finally, the analysis of a 7-year observation campaign of the open cluster NGC 3766 led to the discovery
of a large number of variable stars (Mowlavi et al. \citep{2013A&A...554A.108M}). Amongst them the authors found 36 new variables, which according to the position on the color-magnitude diagram,
have the spectral types between late type B and early type A. The periods of their light variability are from the range of about 0.1 - 0.7 [d],
corresponding to the frequencies of about 1.4 - 10 [d$^{-1}$], and the photometric amplitudes are of the order of a few millimagnitudes.
The very recent work by Mowlavi et al. \cite{2016arXiv161001077M} showed that most of them rotate fast (with $V_{\rm rot}\sin i >150 \kmps$)
which supports the hypothesis that Maia pulsators are fast rotating SPB stars (Salmon et al. \citep{2014A&A...569A..18S}.)

This new type of variables was also suggested from the analysis of space data.
From the Kepler photometry, Balona et al. \cite{2011MNRAS.413.2403B,2015MNRAS.451.1445B,2016MNRAS.460.1318B} found B-type stars with high frequencies and low amplitudes.
These stars are located below the red edge of the $\beta$ Cep instability strip and some of them can be considered as Maia candidates.
Several candidates are also from the CoRoT observations (Degroote et al. \citep{2009A&A...506..471D}).

In this paper, assuming the pulsational origin of this variability, we test a few hypotheses.
Because NGC3766 is a very young open cluster, we consider only Zero Age Main Sequence (ZAMS) models.
The first and already studied hypothesis involves the very high rotation.
In the second, new scenario, we modify the opacity profile in order to excite pulsational modes
in the required frequency range in the rotating models with masses below $M\approx 3.0~M_\odot$.
The last hypothesis demands the modifications of the opacity profile and higher degree modes (up to 6-8)
and works for both, low and fast rotating stars of B8-A2 spectral type.
Sect.\,2 presents the results of our modelling and conclusions are give in Summary.

\section{The three explanations}
\label{sec-1}

{\bf The first hypothesis:\\
fast rotating stars with underestimated masses}

In this scenario, Maia stars are linked to fast rotating Slowly Pulsating B-type stars.
The fast rotation causes that masses of the stars seen close equator-on  are underestimated (Salmon et al. \citep{2014A&A...569A..18S}).
We perform pulsational computations for models with masses $M=3-4~ M_\odot$ in the framework of the traditional approximation (e.g. Lee \& Saio \citep{1997ApJ...491..839L}, Townsend \citep{2003MNRAS.340.1020T}, Savonije \citep{2005A&A...443..557S}).
As mentioned above, NGC3766 is a very young open cluster, thus we analysed only ZAMS models. The initial hydrogen abundance and metallicity were set at 0.7 and 0.015, respectively,
and the chemical mixture was adopted from Asplund at al. \cite{AGSS09}. Here, we show results obtained with the OPAL opacities (Iglesias \& Rogers \citep{OPAL}) but with the OPLIB (Colgan et al. \citep{OPLIB2,OPLIB1})
and OP data (Seaton \citep{OP}), the results are qualitatively similar.

High order gravity modes as well as mixed gravity-Rossby modes are considered. In models with masses corresponding to B8-A2 type, these modes are stable. Their instability begins from the mass of about $M=3 M_\odot$. Moreover, including the effects of rotation on g modes, both, enhances the pulsational instability and causes the shift of modes to the higher frequency range.
In Fig.\,1, we plot the instability parameter, $\eta$, as a function of mode frequency for a ZAMS model with a mass $M=3.2~M_\odot$, effective temperature, $\log T_{\rm eff}=4.119$.
and rotational velocity $V_{\rm rot}=250~\kmps$. The parameter $\eta$ is the normalized work integral and for unstable modes it takes positive values. For clarity, we depicted only modes which become unstable ($\eta >0$) in some range of frequencies.
\begin{figure*}
\centering
\includegraphics[width=1.9\columnwidth,clip]{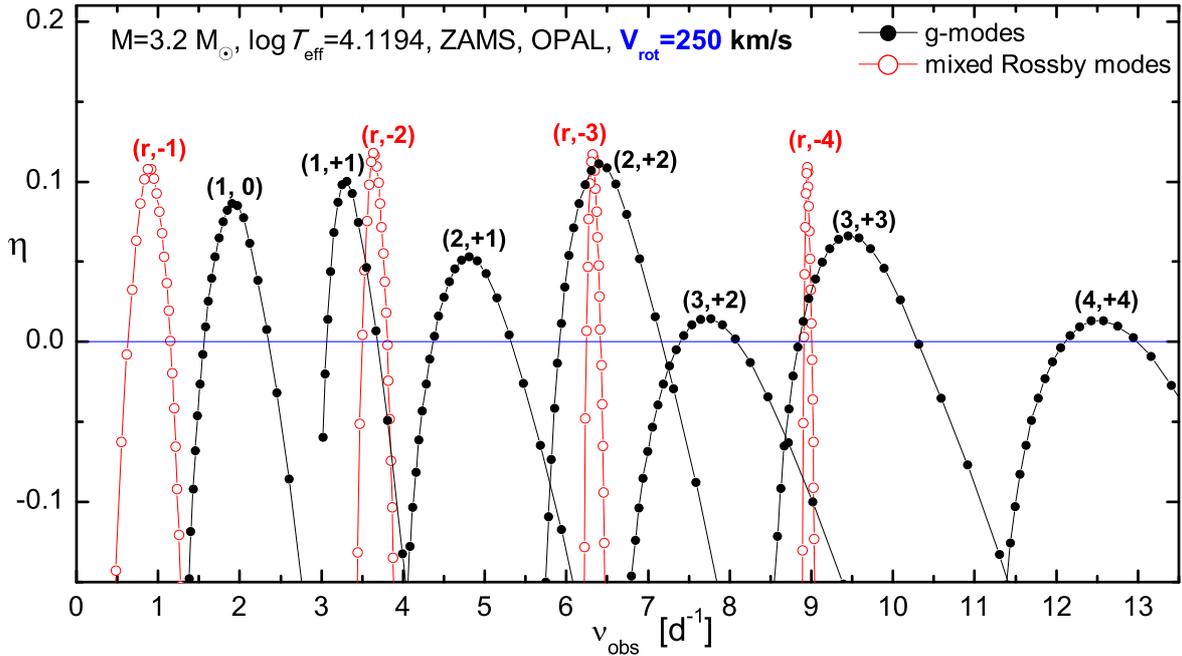}
\caption{The instability parameter, $\eta$, as a function of the mode frequency for a ZAMS model with a mass $M=3.2~M_\odot$
and effective temperature, $\log T_{\rm eff}=4.1194$, rotating with the velocity of $V_{\rm rot}=250~\kmps$.
Only $(\ell,~m)$ and $(r,~-m)$ modes which reach instability ($\eta>0$) are shown.}
\label{fig-1}       
\end{figure*}

{\bf The second hypothesis: \\
fast rotating stars with non-standard opacities}

In the second, new scenario, we continued pulsational computations with the traditional approximation but additionally we modified the mean opacity profile
in models with masses below $M=3.0~M_\odot$. The standard opacities were increased at the depth corresponding to temperature $\log T=5.1$.
This modification mimics the new opacity bump identified in the Kurucz atmosphere models (Cugier \citep{2014A&A...565A..76C}). This new bump was suggested, for example,
as a possible cause of excitation of low frequency g-modes in $\delta$ Scuti stars as detected in the Kepler data (Balona et al. \citep{2015MNRAS.452.3073B}).
The effect of increasing the OPAL opacities by 100\% at $\log T=5.1$ is presented in Fig.\,2. A model with the mass $M=2.5~M_\odot$ on ZAMS was considered.
The rotational rate was $V_{\rm rot}=200~\kmps$.  For models with such masses, pulsational computations with the standard opacities model do not predict any instability.
As one can see, it is more than enough to consider modes with the degree $\ell\le 3$ to cover the frequency range observed in the NGC3766 stars.
\begin{figure*}
\centering
\includegraphics[width=1.9\columnwidth,clip]{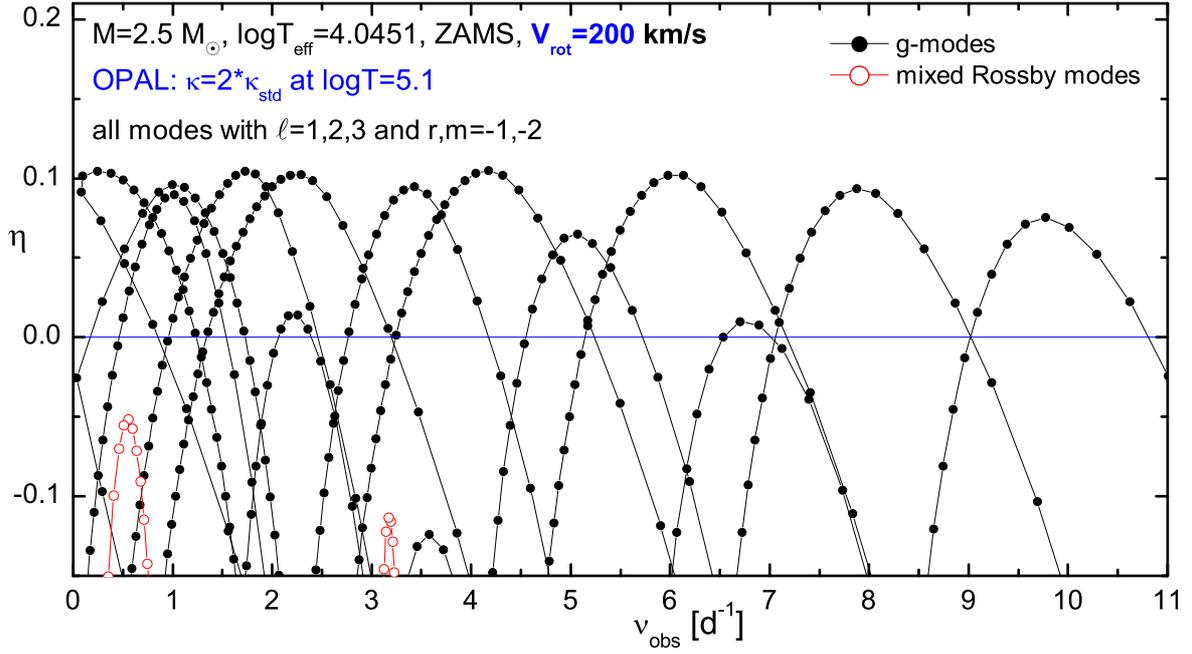}
\caption{The instability parameter, $\eta$, as a function of the mode frequency for the ZAMS model with a mass $M=2.5~M_\odot$
and effective temperature, $\log T_{\rm eff}=4.0451$, rotating with the velocity of $V_{\rm rot}=200~\kmps$.
The model was computed with the modified OPAL opacities which were artificially increased by 100\% at $\log T=5.1$.
Modes with the degree up to 3 and mixed-Rossby modes with $m=-1,~-2$ are shown.}
\label{fig-2}       
\end{figure*}

{\bf The third hypothesis: \\
slow rotating stars with non-standard opacities}

The third explanation involves modifications of the opacity profile and including modes with the harmonic degrees up to 8.
Here, we performed all pulsational computations assuming the zero-rotation approximation, i.e., all effects of rotation on pulsations were neglected.

We found that by increasing the mean opacity at the depth $\log T=5.1$ in a $M=2.5~M_\odot$ model, it is possible to get the instability
in the observed frequency range for modes with $\ell\ge 2$. Adding opacity at the Z-bump ($\log T= 5.3$) makes also the dipole modes unstable.
This scenario works for, both, slow and fast rotating stars of B8-A2 type. Below, we show the results obtained with the modified OPAL data.
We added 150\% of the opacity at $\log T=5.1$ and 50\% of the opacity at $\log T=5.3$.
As one can see, in the mass range corresponding to late B-type and early A-type stars, $M\in(2.5,~4)$, high degree modes ($\ell>6$)
can reach frequencies as high as about 9 d$^{-1}$. Considering modes with such high degrees is justified by the fact that the amplitudes
of the new variable stars in NGC3766 are very low and only very long campaign (about 7 years) allowed to detect them.
\begin{figure*}
\centering
\includegraphics[width=2.0\columnwidth,height=10cm,clip]{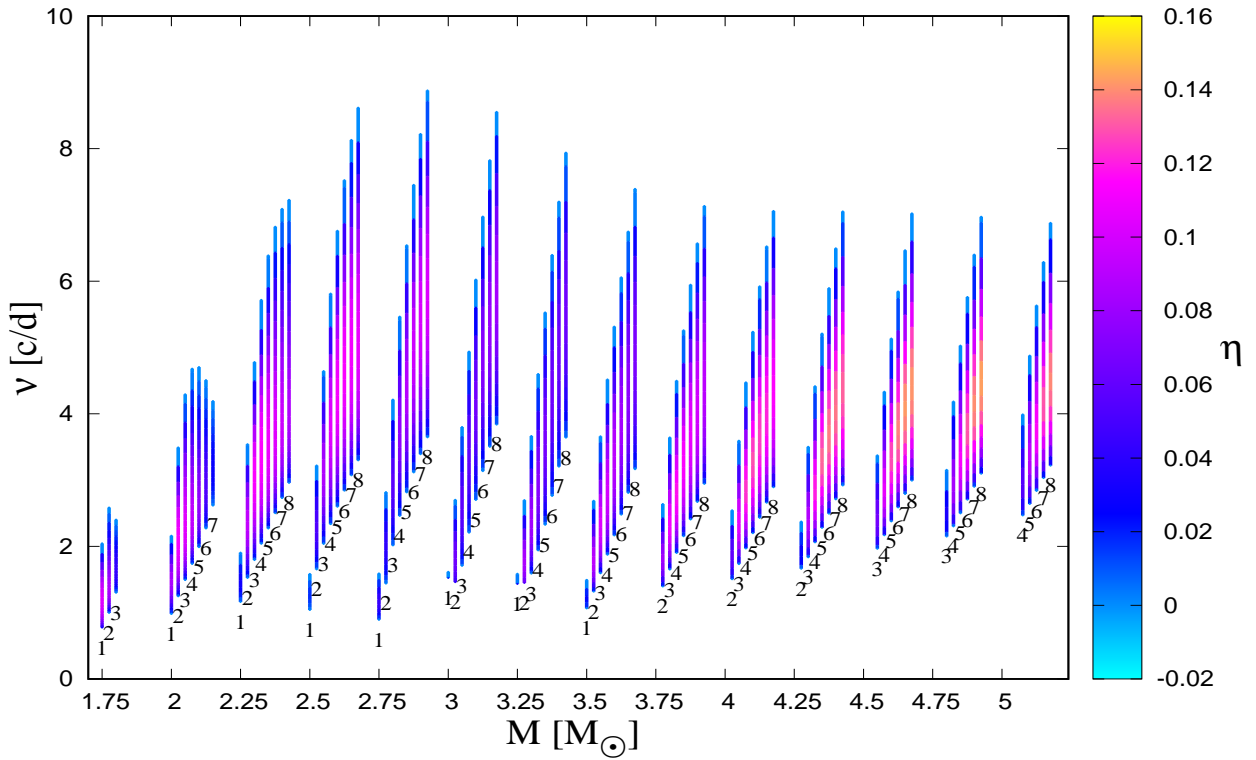}
\caption{The frequency range of unstable modes as a function of mass for ZAMS models computed with the modified OPAL opacity profile.
Modes with the harmonic degrees up to $\ell=8$ were considered. The opacities were increases  by 150\% at $\log T=5.1$ and by 50\% at $\log T=5.3$.
The values of the instability parameter are coded by colours.}
\label{fig-3}       
\end{figure*}

\section{Summary}
\label{sec-con}
Here, we discussed three possible explanations for pulsations in Maia stars. Each of them has its strengths and weaknesses.

Fast rotating stars seen equator-on can account for pulsations of the Maia type. This hypothesis has been already explored by Salmon et al. \cite{2014A&A...569A..18S}.
In this case, we demand the close equator-on orientation for all stars. As a consequence, all modes which have an equatorial node are not good candidates.
It is worth to mention that prograde sectoral modes reach the highest visibility on the equator  (e.g. Salmon et al. \cite{2014A&A...569A..18S}, Daszy\'nska-Daszkiewicz et al. \cite{2007AcA....57...11D}).
On the other hand, given the fact that there are not many identified Maia variables this scenario is plausible.
Moreover, it is strengthened by the recent result of Mowlavi et al. \cite{2016arXiv161001077M} who showed that all new-class variables in NGC 3766 are fast rotators.

Two new scenarios involve modifications of the mean opacity profile. We showed that an increase of the opacity at $\log T=5.1$ is indispensable
to get the instability in the frequency range 1.4 – 10 d$^{-1}$ in the models with masses $M<3.0~M_\odot$.
The above conclusions are valid if the effects of rotation on pulsations are taken in to account.
In the case of g-mode pulsations, we do this via the traditional approximation.
If the effects of rotation on pulsation are neglected then, additionally to the opacity modifications, the higher $\ell$ modes have to be considered.

Here, we need quite significant modifications of the opacities which have to be justified at some point.
There are some results that supports such approach, e.g., by increasing the opacity at the depth $\log T=5.1$
it is possible to explain low frequency g-modes in $\delta$ Scuti stars detected in the Kepler data (Balona et al. \citep{2015MNRAS.452.3073B}).
Given that there are still many uncertainties in computations of stellar opacities and, for example, in case of hybrid B-type pulsators
significant modifications have been suggested (Daszy\'nska-Daszkiewicz et al. \citep{JDD_PW2016}), such hypothesis for Maia variables is reasonable.
Additionally, these two hypothesis do not have any requirements for the inclination angle and all unstable modes can be considered.
The third explanation needs the higher degree modes, up to 8, which have poorer visibility. However, given the low values of the observed amplitudes
and the lack of knowledge of the selection mechanism, we should allow such a possibility.
Moreover, the often forgotten fact is that while the mode visibility is getting worse with higher $\ell$ according to the disc averaging effects, it increases as $\ell^2$
according to the geometrical factor $(\ell-1)(\ell+2)$ (e.g. Daszy\'nska-Daszkiewicz et al. \citep{2002A&A...392..151D}).
The explanation with the opacity modifications and higher $\ell$ modes works for both fast and slow rotators.

To decide which scenario is preferred, firstly, more determinations of the rotational velocity of these stars are needed.
Then, pulsational analysis of more Maia stars has to be performed taking into account various effects on pulsational instability.
It cannot be ruled out that rotation and opacity modification can work together and hold the key to solving the Maia star problem.

\section*{Acknowledgements}
This work was financially supported by the Polish National Science Centre grant 2015/17/B/ST9/02082.
Calculations have been carried out using
resources provided by Wroc{\l}aw Centre for Networking and Supercomputing (http://wcss.pl), grant no. 265.\\

%
%
















%
%


\bibliographystyle{woc}
\bibliography{Biblio}

\end{document}